\documentclass[aps,prd,twocolumn,superscriptaddress,amssymb,eqsecnum,showpacs,showkeyes,secnumarabic,graphics,floatfix,nofootinbib,tightenlines,longbibliography]{revtex4-1}
\usepackage{graphicx}
\usepackage{bm}
\usepackage{dcolumn}
\usepackage{amsmath}
\usepackage{verbatim}
\usepackage{amssymb}
\usepackage{cancel}
\usepackage{hhline}
\usepackage[utf8]{inputenc}
\usepackage[breaklinks=true,colorlinks=true,
linkcolor=blue,urlcolor=blue,citecolor=cyan,
bookmarks=true,bookmarksopenlevel=2]{hyperref}

\numberwithin{equation}{section}
\usepackage{epsfig}

\def \lleq {\lower0.9ex\hbox{ $\buildrel < \over \sim$} ~}
\def \ggeq {\lower0.9ex\hbox{ $\buildrel > \over \sim$} ~}

\def \beq  {\begin{equation}}
\def \eeq  {\end{equation}}
\def \ber  {\begin{eqnarray}}
\def \eer  {\end{eqnarray}}

\usepackage[usenames]{color}

\newcommand{\newc}{\newcommand}

\newc{\diag}{\mathop{\mathrm{diag}}}
\newc{\bear}{\begin{eqnarray}}
\newc{\eear}{\end{eqnarray}}
\newc{\be}{\begin{equation}}
\newc{\ee}{\end{equation}}
\newc{\ba}{\begin{eqnarray}}
\newc{\ea}{\end{eqnarray}}
\newc{\bea}{\begin{eqnarray*}}
\newc{\eea}{\end{eqnarray*}}
\newc{\D}{\partial}
\newc{\ie}{{\it i.e.} }
\newc{\eg}{{\it e.g.} }
\newc{\etc}{{\it etc.} }
\newc{\etal}{{\it et al.}}
\newc{\lcdm }{$\Lambda$CDM }
\newcommand{\nn}{\nonumber}
\newc{\ra}{\rightarrow}
\newc{\lra}{\leftrightarrow}
\newc{\lsim}{\buildrel{<}\over{\sim}}
\newc{\gsim}{\buildrel{>}\over{\sim}}

\newcommand{\mincir}{\raise
-3.truept\hbox{\rlap{\hbox{$\sim$}}\raise4.truept\hbox{$<$}\ }}
\newcommand{\magcir}{\raise
-3.truept\hbox{\rlap{\hbox{$\sim$}}\raise4.truept\hbox{$>$}\ }}

\begin{document}

\title{Geodesics of McVittie Spacetime with a Phantom Cosmological Background}

\author{Ioannis Antoniou}\email{ianton@cc.uoi.gr}
\author{Leandros Perivolaropoulos}\email{leandros@uoi.gr}
\affiliation{Department of Physics, University of Ioannina,Greece}

\date{\today}

\begin{abstract}
We investigate  the geodesics of a Schwarzschild spacetime
embedded in an isotropic expanding cosmological background
(McVittie metric). We focus on bound particle geodesics in  a
background including matter and phantom dark energy with constant
dark energy equation of state parameter $w<-1$ involving a future
Big Rip singularity at a time $t_{\ast}$. Such geodesics have been
previously studied in the Newtonian approximation and found to
lead to dissociation of bound systems at a  time
$t_{rip}<t_{\ast}$ which for fixed background $w$, depends on a single dimensionless
parameter $\bar{\omega}_0$ related to the angular momentum and
depending on the mass and the size of the bound system. We extend
this analysis to large massive bound systems where the
Newtonian approximation is not appropriate and we compare the
derived dissociation time with the corresponding time in the
context of the Newtonian approximation. By identifying the time
when the effective potential minimum disappears due to the
repulsive force of dark energy we find that the dissociation time
of bound systems occurs earlier than the prediction of the
Newtonian approximation. However, the effect is negligible for all
existing cosmological bound systems and it would become important
only in hypothetical bound extremely massive ($10^{20} M_\odot$)
and large ($100Mpc$) bound systems. We verify this result by
explicit solution of the geodesic equations. This result is due to
an interplay between the repulsive phantom dark energy effects and
the existence of the well known innermost stable orbits of
Schwarzschild spacetimes.

\end{abstract}
\pacs{98.80.-k,04.25.-g,98.80.Bp}

\maketitle

\section{Introduction}
The simplest cosmological model that is consistent with current
cosmological observations is the $\Lambda$CDM model where the
observed accelerating expansion of the universe is attributed to a
cosmological constant which introduces repulsive properties to
gravity at large distances
\cite{Perivolaropoulos:2006ce,Carroll:2000fy,Padmanabhan:2002ji,Peebles:2002gy,Copeland:2006wr,Frieman:2008sn}.
The cosmological constant may be described as a homogeneous dark
energy perfect fluid with constant energy density and negative
pressure with constant equation of state \be w=\frac{p}{\rho}=-1
\label{stateeq}\ee A generalization of $\Lambda$CDM where the
cosmic acceleration is induced by a dark energy fluid with
constant equation of state introduces a new parameter $w$ in the
models which is constrained by cosmological observations at the $1
\sigma$ level to be in the range
\cite{Said:2013jxa,Ade:2015rim,Zheng:2014ara,Piedipalumbo:2013dqa}
 \be -1.5<w<-0.7 \ee Based on these
constraints and in the context of the above minimal generalization
of $\Lambda$CDM there is a significant probability that $w<-1$.
For such a range of $w$, this class of models predicts the
existence of a future singularity where the scale factor diverges
at a finite future time.

This behavior emerges by solving the Friedmann equation in the
presence of matter density $\rho_m$ and dark energy density
$\rho_x$ which may be written as
~\cite{Johri:2003rh},~\cite{Nesseris:2004uj}
 \be \frac{\dot{a}^{2}}{a^{2}}=\frac{8\pi G}{3}[\rho_{m}+\rho_{x}]=
 H_{0}^{2}[\Omega_{m}^{0}(\frac{a_{0}}{a})^{3}+\Omega_{x}^{0}(\frac{a_{0}}{a})^{3(1+w)}] \ee

 and \bear \frac{\ddot{a}}{a}&=&-\frac{4\pi
 G}{3}[\rho_{m}+\rho_{x}(1+3w)]=-\frac{4\pi
 G}{3}\rho_{x}[\Omega_{x}^{-1}+3w] \nn\\ & & =-\frac{4\pi
 G}{3}\rho_{x}[\frac{\Omega_{m}^{0}}{\Omega_{x}^{0}}(\frac{a_{0}}{a})^{-3w}+1+3w]\eear
 with solution \be
 a(t)=\frac{a(t_{m})}{[-w+(1+w)\frac{t}{t_{m}}]^{-\frac{2}{3(1+w)}}}, \hspace{0.5cm}
    t>t_{m}\label{scalef} \ee
where $t_m$ is the time when the dark energy density becomes
larger than the matter density. For $w<-1$ the scale factor and
its derivatives diverge at a finite time known as the Big Rip time
~\cite{Caldwell:2003vq,Nojiri:2005sx,Chimento:2004ps,Cattoen:2005dx}
\be t_{*}=\frac{w}{1+w}t_{m} > 0 \ee

This divergence results in a diverging repulsive gravitational
force which rips apart all bound systems at times $t_{rip}$ that
depend on their binding energies and forms of effective
potentials.

An important question to address is {\it What is the physical
mechanism that induces this dissociation of bound systems and what
is the time when the dissociation occurs as a function of $w$?} In
order to address this question, a gravitationally bound system may
be represented as a single test particle bound in a circular orbit
of radius $r_0$  by the gravitational force of a central spherical
massive object of mass $m$. The features of the trajectory of the
test particle may be obtained in any of the following ways

\begin{enumerate}
    \item  Using a rough comparison of the attractive gravitational
force with the repulsive force induced by the expansion
~\cite{Caldwell:2003vq}.
    \item  By using  a derivation of the particle trajectory
using equations of motion in the Newtonian approximation which
take into account the attractive gravitational force, the
repulsive force due to the expansion as well as  the centrifugal
effects due to angular momentum ~\cite{Nesseris:2004uj},
\cite{Nandra:2013jga,Gao:2011tq,Nandra:2011ui}.
    \item  Using the full relativistic geodesic equations obtained
     from a metric that is a solution of  the Einstein equations
     and interpolates between a Schwarzschild metric and an FRW
     metric. Such a metric is the McVittie metric
     ~\cite{mcvittie}. Other approaches to such an interpolation
     may be found in Refs ~\cite{Moradi:2015caa,Einstein:1945id,Bona,Baker:2001yc}

\end{enumerate}

Previous studies have pursued the first two approaches with
results that are in qualitative agreement within a factor of 3.
According to the approach of Ref ~\cite{Nesseris:2004uj}, the
dissociation of the bound system is associated with the
disappearance of the minimum of the effective potential that
determines the radial motion of the test particle. This minimum
disappears when the dynamics become dominated by the effects of
the accelerating expansion of the phantom cosmological background.
Thus the dissociation of a bound system occurs at a time $t_{rip}$
given by \be
t_{*}-t_{rip}=\frac{16\sqrt{3}}{9}\frac{T\sqrt{2|1+3w|}}{6\pi|1+w|}\label{tripn}\ee
where $T$ is the period of the gravitationally bound system with mass $m$, radius $r_0$ and
angular velocity $\omega_0$ of the form \be \omega_{0}^{2}\equiv
(\frac{2\pi}{T})^2=\frac{G m}{r_{0}^{3}}\ee

This result improves over the corresponding result of Ref.
~\cite{Caldwell:2003vq} by the factor $16 \sqrt{3}/9\simeq 3$ because it
takes into account the effects of the centrifugal term and
provides a clear definition of the dissociation time as the time
when the minimum of the effective potential disappears due to the
domination of the repulsive gravitational effects of the
expansion. On the other hand, the analysis of
Ref.~\cite{Nesseris:2004uj} is limited by the fact that it uses
the Newtonian approximation for the dynamical equations of the
particle orbits and therefore it may not be applicable for the
analysis of the dissociation of strongly bound systems like
accretion disks ~\cite{Pringle:1981ds},~\cite{Abramowicz:2011xu}.

In this study we extend the analysis of
Ref.~\cite{Nesseris:2004uj} by going beyond the Newtonian
approximation and taking into account relativistic effects. In
particular, we consider the full geodesics corresponding to the
McVittie metric in a phantom cosmological background. Using these geodesic equations we construct the
relativistic effective potential corresponding to bound particle
orbits and derive the time of dissociation ($t_{rip}$) when the
minimum of the potential disappears due to expansion effects.
These results are confirmed by comparing with numerical solutions
of the geodesic equations corresponding to initial circular
bounded orbits. We compare these results with the corresponding
results of previous studies ~\cite{Nesseris:2004uj} obtained in
the Newtonian limit.

 The structure of this paper is the following: In the next
section we review the McVittie metric  and
its limits (FRW, Newtonian, Schwarzschild). We also analyze the form
of the geodesics, define the effective potential that determines
the dynamics of the bound orbits and compare it with the
corresponding Newtonian approximation in the context of a phantom
cosmology. In section 3 we present the numerical solution of the
geodesics for various parameter values showing the dissociation of
the bound systems. The times of dissociation $t_{rip}$ obtained by
the numerical solution are also compared with the time when the
minimum of the effective potential disappears due to the repulsive
effects of the accelerating cosmological expansion. Comparison
with the corresponding Newtonian results is also made. Finally in
section 4 we conclude, summarize and discuss possible extensions
on this analysis.

\section {Geodesic equations and their limits}

An acceptable way to describe a bound system embedded in an
expanding cosmological background is provided by the McVittie
metric ~\cite{mcvittie}.  For a flat cosmological
background this metric is of the form
\be ds^2=-(f-\frac{r^2H^2}{c^2})d(ct)^2-2rHf^{-1/2}dt dr+
f^{-1}dr^2+r^2d\Omega^2\label{metric}\ee where $m>0$ is a constant,
$f=f(r)=1-2Gm/(c^2 r) >0$ and $H=H(t)=\frac{\dot a}{a}$ is the
Hubble parameter of the cosmological background. In what follows
we do not set $c=G=1$ in order to clearly show the Newtonian limit
($c\rightarrow \infty$).

In eq. (\ref{metric}) $r$ is the physical
spatial coordinate connected with  the comoving spatial coordinate
$\rho$ as $\rho=\frac{r}{a(t)}$. Setting $m=0$ and using the
comoving coordinate we obtain the flat background FRW metric

\bear ds^2&=&-(1-r^2H^2)d(ct)^2-2rHdt dr+ dr^2+r^2d\Omega^2 \nn\\
& & =-dt^2+a^2(d\rho^2+\rho^2d\Omega^2)\eear  Similarly, setting
$H=0$ the metric (\ref{metric}) reduces to the Schwarzschild metric.

The Schwarzschild de Sitter metric may also be obtained as a special
case of the McVittie metric by fixing the Hubble parameter to a
constant $H^2=H_0^2=\frac{\Lambda}{3}$ and performing a coordinate
transformation\cite{Nolan:2014maa} \be T=t+u(r)\ee with \be
u'(r)=H_0r/c(\sqrt{f}(f-\frac{r^2H^2}{c^2}))\ee leading to the
Schwarzschild de Sitter (or Kottler) metric \bear
ds^2&=&-(1-\frac{2Gm}{c^2r}-\frac{\Lambda}{3}r^2)d(cT)^2\nn\\ & &
-(1-\frac{2Gm}{c^2r}-\frac{\Lambda}{3}r^2)^{-1}dr^2+r^2d\Omega^2\eear
In the Newtonian limit, using comoving coordinates, the McVittie
metric may me written as
~\cite{Nesseris:2004uj},~\cite{Faraoni:2007es},~\cite{Noerdlinger}
\be
ds^2=(1-\frac{2Gm}{c^2a(t)\rho})d(ct)^2-a(t)^2(d\rho^2+\rho^2(d\theta^2+\sin^2\theta
d\varphi^2))\label{metricn}\ee

The Newtonian geodesics corresponding to the metric
(\ref{metricn}) are of the form
~\cite{Baker:2001yc},~\cite{Price:2005iv} \be
\ddot{r}-\frac{\ddot{a}}{a}r+\frac{Gm}{r^2}-r\dot{\varphi}^2=0\label{ddotr}\ee
and \be r^2\dot{\varphi}=L \label{angular}\ee where $r$ is the
physical coordinate ($r=a \rho$) and $L$ is the angular momentum
per unit mass ($L=\omega r^2$, constant). Combining eqs
(\ref{ddotr}) and (\ref{angular}) we find the radial dynamical
equation in the Newtonian limit \be
\ddot{r}=\frac{\ddot{a}}{a}r+\frac{L^2}{r^3}-\frac{Gm}{r^2}\label{ddotn}\ee
Notice that $c$ does not appear in this equation since it is
non-relativistic. If we ignore the term due to the expansion, then
the angular velocity of a test particle in a bound circular orbit
with radius $r_0$ at an initial time $t_0$ is obtained from
eq. (\ref{ddotn}) as \be
\dot{\varphi}(t_0)^2=\omega_0^2=\frac{Gm}{r_0^3}\label{fdot}\ee The
radius of the circular orbit will be perturbed once the expansion
is turned on but the above eq. (\ref{fdot}) remains a good
approximation close to the end of the era of matter domination
(eq. (\ref{ddotn})) $t_m=t_0$, when the expansion repulsive force is
subdominant. It is convenient to rescale eq. (\ref{ddotn}) to a
dimensionless form by defining the dimensionless quantities
$\overline{r}\equiv\frac{r}{r_0}$, $\overline{\omega}_0\equiv\omega_0
t_0$ and $\overline{t}\equiv\frac{t}{t_0}$. The choice of this
rescaling is made so that the effect of the expansion is initially
small (at time ${\bar t}=1$) and the  initial minimum of the
effective potential is approximately at ${\bar r}=1$. Typical
values of ${\bar \omega}_0$ are obtained using the scale and the
mass of bound systems. Thus ${\bar \omega}_0$ is $O(1)$ for a
cluster of galaxies, about 200 for a galaxy and  $10^6$ for the
solar system.

Assuming a constant $w$ and using the form of the scale factor in
eq. (\ref{scalef}), the radial dynamical equation (\ref{ddotn})
takes the form
\be\ddot{\overline{r}}+\frac{\overline{\omega}_0^2}{\overline{r}^2}(1-\frac{1}{\overline{r}})+\frac{2}{9}\frac{(1+3w)\overline{r}}{(-w+(1+w)\overline{t})^2}=0\label{rfinal}\ee
From eq (\ref{rfinal}) we derive the effective radial force \be
F_{eff}=-\frac{\overline{\omega}_0^2}{\overline{r}^2}(1-\frac{1}{\overline{r}})-\frac{2}{9}\frac{(1+3w)\overline{r}}{(-w+(1+w)\overline{t})^2}\ee
and the corresponding effective potential \be
V_{eff}=-\frac{\overline{\omega}_0^2}{\overline{r}}+\frac{\overline{\omega}_0^2}{2\overline{r}^2}-\frac{1}{2}\lambda(\overline{t})^2\overline{r}^2\ee
where (for $w<-1$) \be
\lambda^2(\overline{t})=\frac{2}{9}\frac{(1+3w)}{(-w+(1+w)\overline{t})^2}\ee The repulsive
term due to the expansion (proportional to $\lambda^2$) increases
with time and at a time ${\bar t}_{rip}$ given by eq. (\ref{tripn}), it
destroys the effective potential minimum induced by the interplay
between the attractive gravity and centrifugal terms. Thus  a
bound system gets dissociated by the expansion at ${\bar t}={\bar t}_{rip}$
~\cite{Nesseris:2004uj}.

This analysis made in the context of the Newtonian approximation
is inappropriate for some massive large strongly bound systems
where relativistic effects need to be taken into account. A proper
relativistic analysis requires the use of the geodesic equations
obtained from the McVittie metric eq. (\ref{metric}). These
dynamical equations are of the form ~\cite{Nolan:2014maa} \be
\ddot{r}=rf^{1/2}H'\dot{t}^2+(1-\frac{3Gm}{c^2r})\frac{L^2}{r^3}-\frac{Gm}{r^2}+rH^2\label{radialeq}\ee
\be
\ddot{t}=-(1-\frac{3Gm}{rc^2})f^{-1/2}H\dot{t}^2-\frac{2Gm}{r^2}f^{-1}\dot{t}\dot{r}+f^{-1/2}H\label{timeeq}\ee
The overdot represents the derivative with respect to the proper
time and the prime represents derivative with respect to the
coordinate time. A first integral of these equations may also be
obtained as \be
\chi\dot{t}^2+2\frac{\alpha\dot{t}\dot{r}}{c}-\frac{f^{-1}\dot{r}^2}{c^2}-\frac{L^2}{c^2r^2}=1 \label{fint}\ee
where \be \chi(t,r)=f-\frac{r^2H^2}{c^2}, \hspace{0.5cm}
\alpha(t,r)=\frac{rf^{-1/2}H}{c}\ee We may choose ${\dot t} >0$
along causal geodesics and focus on the system of the radial
geodesic eq. (\ref{radialeq}) coupled with the first integral (\ref{fint}).

\begin{figure*}[!t]
\centering
\rotatebox{0}{\resizebox{0.48\textwidth}{!}{\includegraphics{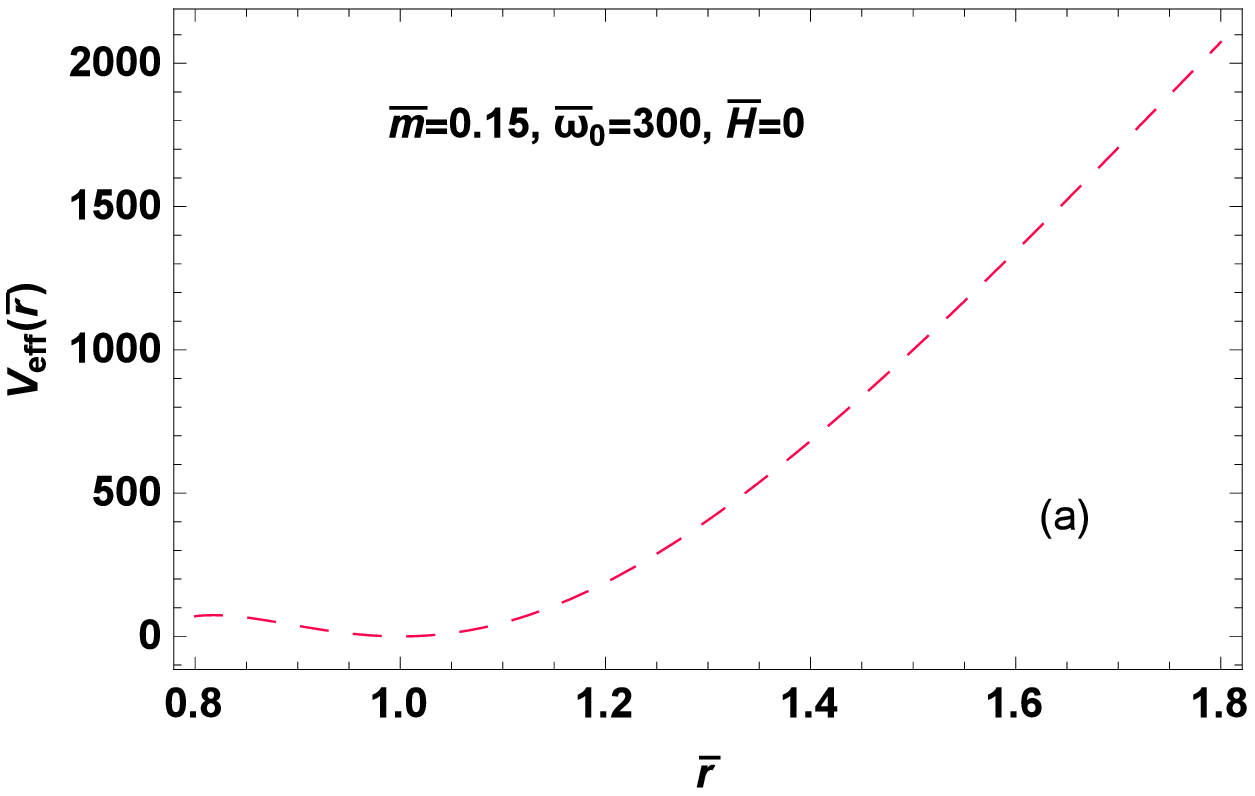}}}
\rotatebox{0}{\resizebox{0.48\textwidth}{!}{\includegraphics{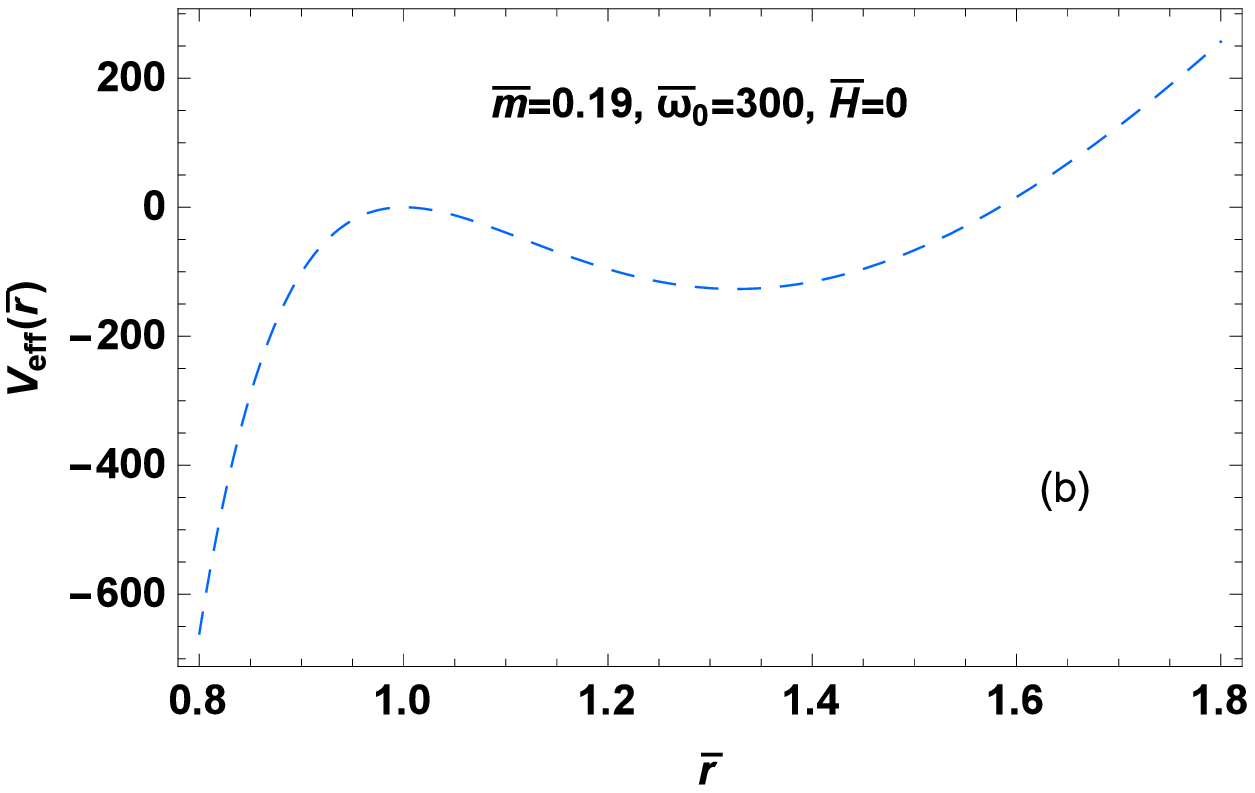}}}
\hspace{0pt}\caption{ The effective potential as a function of
$\bar r$ in a static universe when $\bar{\omega}_0=300$ for $\bar
m=0.15<\frac{1}{6}$ and $\bar m=0.19>\frac{1}{6}$.}\vspace{1cm} \label{figa}
\end{figure*}

\begin{figure}[!ht]
\centering
\includegraphics[scale=0.6]{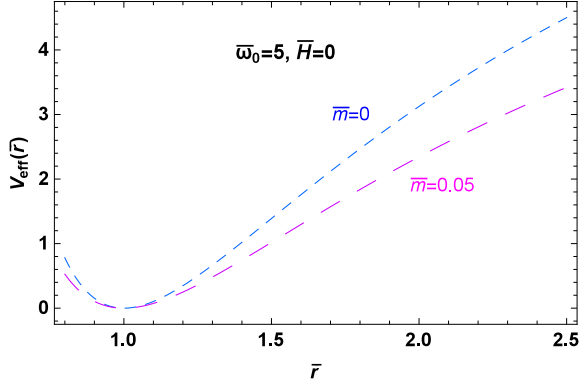}
\caption{The effective potential for ${\bar m}=0$ and ${\bar
m}=0.05$ with the effects of expansion turned off when
$\bar{\omega}_0=5$} \label{fig:plotvng1}
\end{figure}

As a first step towards the investigation of this system we use a proper
rescaling. In
particular we assume a background expansion model corresponding to
constant $w<-1$ (eq. (\ref{scalef})) and rescale the system using the
scales $r_0$ (circular orbit radius in the absence of expansion)
and $t_0=t_m$. We then define the dimensionless quantities: $\bar
t\equiv t/t_0$, $\bar \tau\equiv \tau/t_0$ ($\tau$ is the proper time), $\bar r\equiv r/r_0$,
$\bar m\equiv Gm/r_0c^2$, $\bar H\equiv Ht_0$, $\bar
\omega_0\equiv\omega_0 t_0$. Using the dimensionless coordinates,
the radial geodesic (\ref{radialeq}) and the first integral (\ref{fint}) take the
form \be \ddot{\bar r}=\bar r f^{1/2}\bar H'\dot{\bar t}^2
+(1-\frac{3\bar m}{\bar r})\frac{\bar \omega_0^2}{\bar r^3}-
\frac{\bar m}{\bar r^2}(\frac{ct_0}{r_0})^2+\bar r\bar H^2
\label{radialeq1}\ee

\bear [ f-(\frac{r_0}{ct_0})^2\bar r^2\bar H^2]\dot{\bar
t}^2+2(\frac{r_0}{ct_0})^2\bar r\bar H f^{-1/2}\dot{\bar t}\dot{\bar r}\nn\\
-\frac{\dot {\bar r}^2}{ f} (\frac{r_0}{ct_0})^2- \frac{\bar
\omega_0^2}{\bar r^2}(\frac{r_0}{ct_0})^2=1 \eear
where $f$ is expressed in terms of $\bar m$ as
\be
f=1-\frac{2 {\bar m}}{\bar r} \label{barf}
\ee
We now determine the scale $r_0$ for the relativistic case
considered here and compare with the corresponding Newtonian
scale. The effective radial force in the absence of cosmological
expansion ($H=0$) takes the form \be F_{eff}=(1-\frac{3\bar
m}{\bar r})\frac{\bar \omega_0^2}{\bar r^3}- \frac{\bar m}{\bar
r^2}(\frac{ct_0}{r_0})^2\label{eff}\ee which vanishes for ($\bar
r=1$) \be {\bar \omega_0}=\frac{ct_0}{r_0}\sqrt{\frac{\bar
m}{1-3\bar m}}\label{om0}\ee

eq. (\ref{om0}) constitutes also the definition of the scale $r_0$
used for the rescaling of the geodesic equations. From
eqs (\ref{radialeq}) and (\ref{om0}) we obtain the dimensionless
form of the radial geodesic equation \be \ddot{\bar r}=\bar r
f^{1/2}\bar H'\dot{\bar t}^2 +(1-\frac{3\bar m}{\bar r})\frac{\bar
\omega_0^2}{\bar r^3}- \frac{(1-3\bar m)\bar \omega_0^2}{\bar
r^2}+\bar r\bar H^2\label{fradialeq}\ee Similarly, the
dimensionless form of the first integral eq. (\ref{fint}) is \bear
[f-\frac{\bar r^2\bar m\bar H^2}{\bar \omega_0^2(1-3\bar m)}]\dot{\bar t}^2+
\frac{2\bar m}{\bar \omega_0^2(1-3\bar m)}\bar r\bar H f^{-1/2}\dot{\bar t}\dot{\bar r}\nn\\
-\frac{\dot{\bar r}^2}{ f\bar \omega_0^2}\frac{\bar m}{1-3\bar
m}-\frac{\bar m}{\bar r^2(1-3\bar m)}=1 \label{ftimeeq}\eear

\begin{figure*}[!t]
\centering
\rotatebox{0}{\resizebox{0.48\textwidth}{!}{\includegraphics{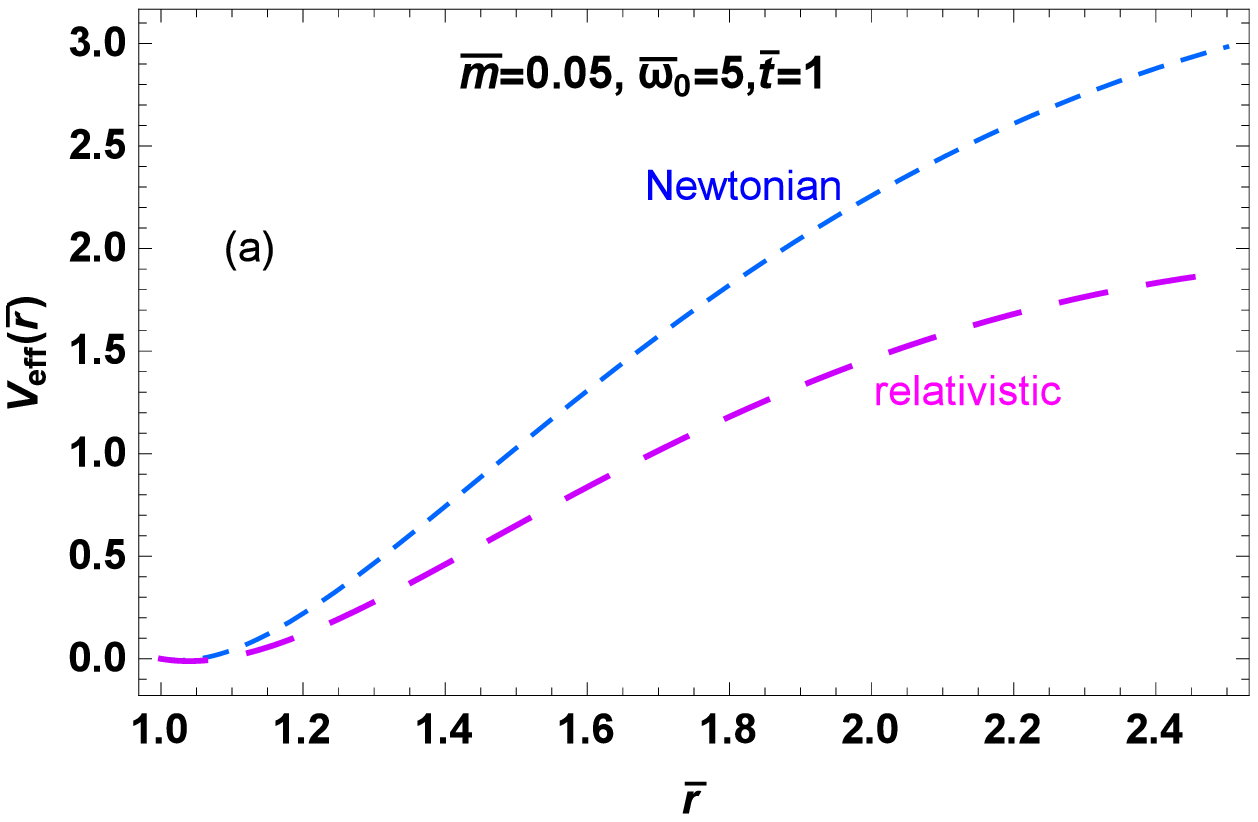}}}
\rotatebox{0}{\resizebox{0.48\textwidth}{!}{\includegraphics{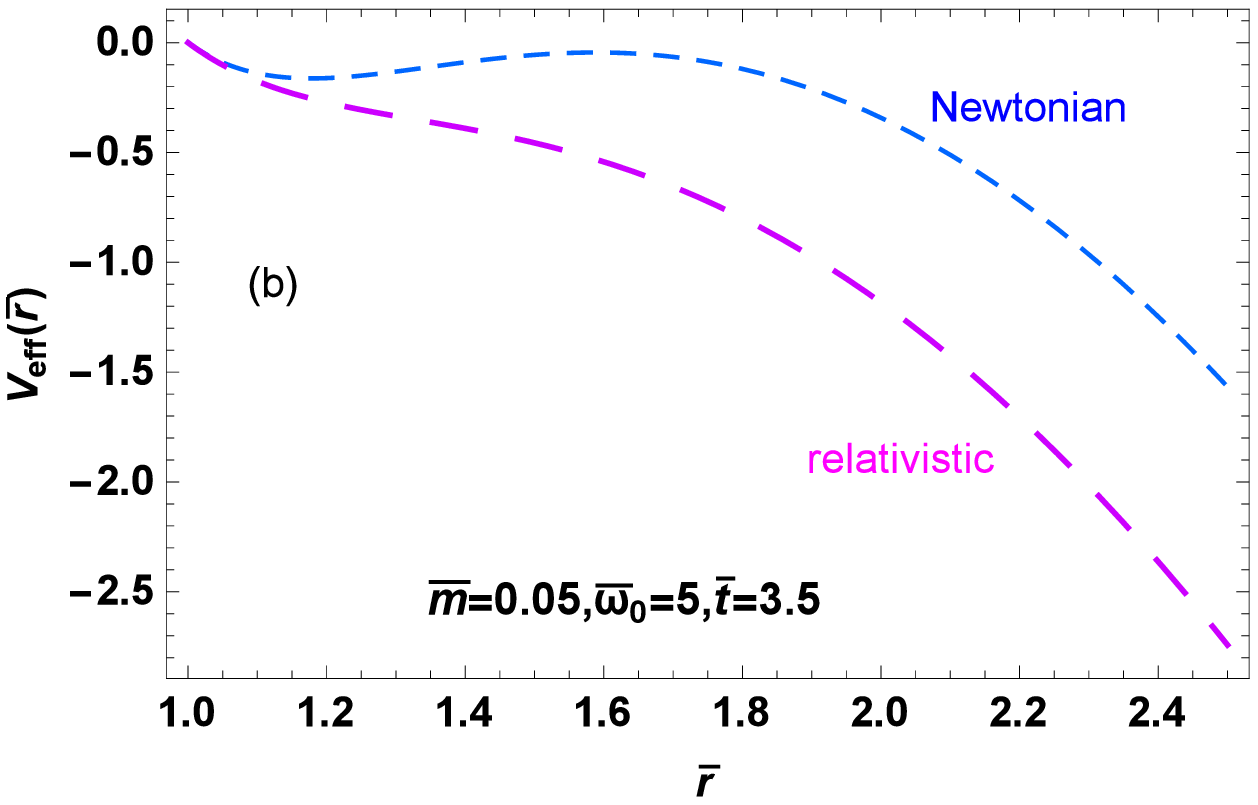}}}
\hspace{0pt}\caption{Fig. 3a: The effective potential with the
effects of the expansion have been turned on ($H\neq 0$, $w=-1.2$)
but the time shown is before the bound system dissociation time
${\bar t}_{rip}$. Fig. 3b: The form of the effective potential for $t=3.5
t_m$, where the system has been dissociated according to the full
relativistic analysis but it remains bound according to the
Newtonian approximation.}\vspace{1cm} \label{figb}
\end{figure*}

The Newtonian limit is obtained for $c\rightarrow \infty$ which
corresponds to
 \be{\bar m}\equiv \frac{G m}{c^2 r_0}  \rightarrow 0, \hspace{0.5cm} f \rightarrow 1  \ee As expected
in this limit we obtain ${\dot {\bar  t}}=1$ from the integral equation
(\ref{ftimeeq}) while the radial equation reduces to the
corresponding Newtonian equation (\ref{ddotn}). Similarly, in this
limit the scale $r_0$ (defined through (\ref{om0})) reduces to the corresponding Newtonian scale
(eq. (\ref{fdot})) since $c^2 {\bar m}=\frac{G m}{r_0}$.

Therefore, assuming a fixed expanding cosmological background, the
geodesics in the McVittie metric are fully determined by two
dimensionless parameters $\bar m$ and $\bar \omega_0$ while the
corresponding Newtonian orbits are determined by a single
parameter ($\bar \omega_0$) and are obtained as the limit ${\bar
m}\rightarrow 0$ of the relativistic orbits. The dimensionless
parameters $\bar m$ and $\bar \omega_0$ are obtained from the mass
$m$ (measured in solar masses $M_\odot$) and the scale $r_0$
(measured in Mpc) of the physical system by the relations \be \bar
m\simeq \frac{5\times10^{-20}m}{{r_0}}\label{mbarvalue}\ee \be
\bar \omega_0\simeq \frac{1780}{r_0}\sqrt{\frac{\bar m}{1-3\bar
m}}\label{om0value}\ee while the reverse relations are \be m\simeq
\frac{3.5\times10^{22}\bar m}{\bar \omega_0}\sqrt{\frac{\bar
m}{1-3\bar m}}\ee \be r_0\simeq \frac{1780}{\bar
\omega_0}\sqrt{\frac{\bar m}{1-3\bar m}}\ee In the Schwarzschild
limit ($H=0$) the radial geodesic equation becomes \be \ddot{\bar
r}=(1-\frac{3\bar m}{\bar r})\frac{\bar \omega_0^2}{\bar r^3}-
\frac{(1-3\bar m)\bar \omega_0^2}{\bar r^2}\label{fradq}\ee The
effective radial force (RHS) has two roots given by \be \bar r=1, \hspace{0.5cm}
   \bar r=\frac{3\bar m}{1-3\bar m}\ee

The root  ${\bar r}=1$ is easily shown (by considering the
derivative of the effective force) to correspond to  a stable
circular orbit for ${\bar m}< \frac{1}{6}$ while for
$\frac{1}{6}<\bar m<\frac{1}{3}$ the root  ${\bar r}=\frac{3\bar
m}{1-3\bar m}>1$ corresponds to a (weakly) stable circular orbit.
We therefore recover the well known fact that the innermost stable
circular orbit of the Schwarzschild metric is obtained for $\bar m
=\frac{1}{6}$ which corresponds to a radius $r_0=\frac{6 G
m}{c^2}$.

In Fig. 1 we show the effective potential obtained by integration
of the effective force of eq. (\ref{eff}) for ${\bar
m}=0.15<\frac{1}{6}$ and for ${\bar m}=0.19>\frac{1}{6}$.

The plot shows the development of the local maximum of the
effective potential at ${\bar r}=1$ when ${\bar m}>\frac{1}{6}$
and the development of a new minimum at ${\bar r}>1$.
Interestingly, the new minimum is weaker and there is less
restoring force for perturbations towards larger ${\bar r}$. Thus,
as ${\bar m}$ increases towards the limiting value of
$\frac{1}{3}$ (beyond this value there is no circular orbit) the
circular orbit becomes less stable and susceptible to
destabilization by the repulsive effects of the accelerating
expansion.

We now turn on the expansion to investigate how this affects the
effective radial force and the potential of the radial geodesics.
For definiteness we set $w=-1.2$ (${\bar t}_*=6$) which corresponds to a phantom
background expansion consistent with current observational
constraints ~\cite{Said:2013jxa}. The effective force may be
obtained in the general relativistic geodesics  when expansion is
present by solving the first integral eq. (\ref{ftimeeq}) for
${\dot {\bar t}}^2$ and substituting in the radial geodesic
eq. (\ref{fradialeq}). Assuming a slow shift of the location of the
potential minimum with time we ignore the terms proportional to
$\dot {\bar r}$ in constructing the effective force and the effective
potential. This approximation is justified in the next section
where we obtain the numerical solution of the full system of the
coupled geodesic equations (\ref{ftimeeq}) and (\ref{fradialeq}). The effective force thus obtained is of
the form \bear F_{eff}&=&\bar r f^{1/2}\bar H'[\frac{1+\frac{\bar
m}{\bar r^2(1-3\bar m)}} { f-\frac{\bar r^2{\bar H}^2\bar m}{\bar
\omega_0^2(1-3\bar m)}}] \nn\\& &+(1-\frac{3\bar m}{\bar
r})\frac{\bar \omega_0^2}{\bar r^3}- \frac{(1-3\bar m)\bar
\omega_0^2}{\bar r^2}+\bar r\bar {H}^2\label{ffrad}\eear

The corresponding effective potential may be obtained by
integrating numerically the effective force $F_{eff}$ as \be
V_{eff}({\bar r})=-\int_1^r F_{eff}(\bar r') d\bar r' \label{veff}\ee

In Fig. 2 we show a plot of the effective potential for ${\bar
m}=0$ and ${\bar m}=0.05$ with the effects of expansion turned
off. The plot shows that the relativistic effects tend to make the
bound state weaker and more susceptible to dissociation due to the
effects of the expansion. This effect is related to the
development of the local maximum (Fig. 1) of the relativistic
potential for a radius smaller than the radius of the stable orbit
(potential minimum) which is also the reason for the existence of
an innermost stable circular orbit. Thus in contrast to naive
intuition, the stronger effects of gravity in the relativistic
case tend to destabilize rather than stabilize  bound systems.

This is also demonstrated in Fig. 3a where the effects of the
expansion have been turned on ($H\neq 0$, $w=-1.2$) but the time
shown is before the bound system dissociation time ${\bar t}_{rip}$.
Clearly, the binding power of the potential has been weakened on
large scales in both the relativistic (lower curve) and the
Newtonian case (upper curve). Fig. 3b shows the form of the
effective potential for $t=3.5 t_m$.

\begin{figure*}[!t]
\centering
\rotatebox{0}{\resizebox{0.48\textwidth}{!}{\includegraphics{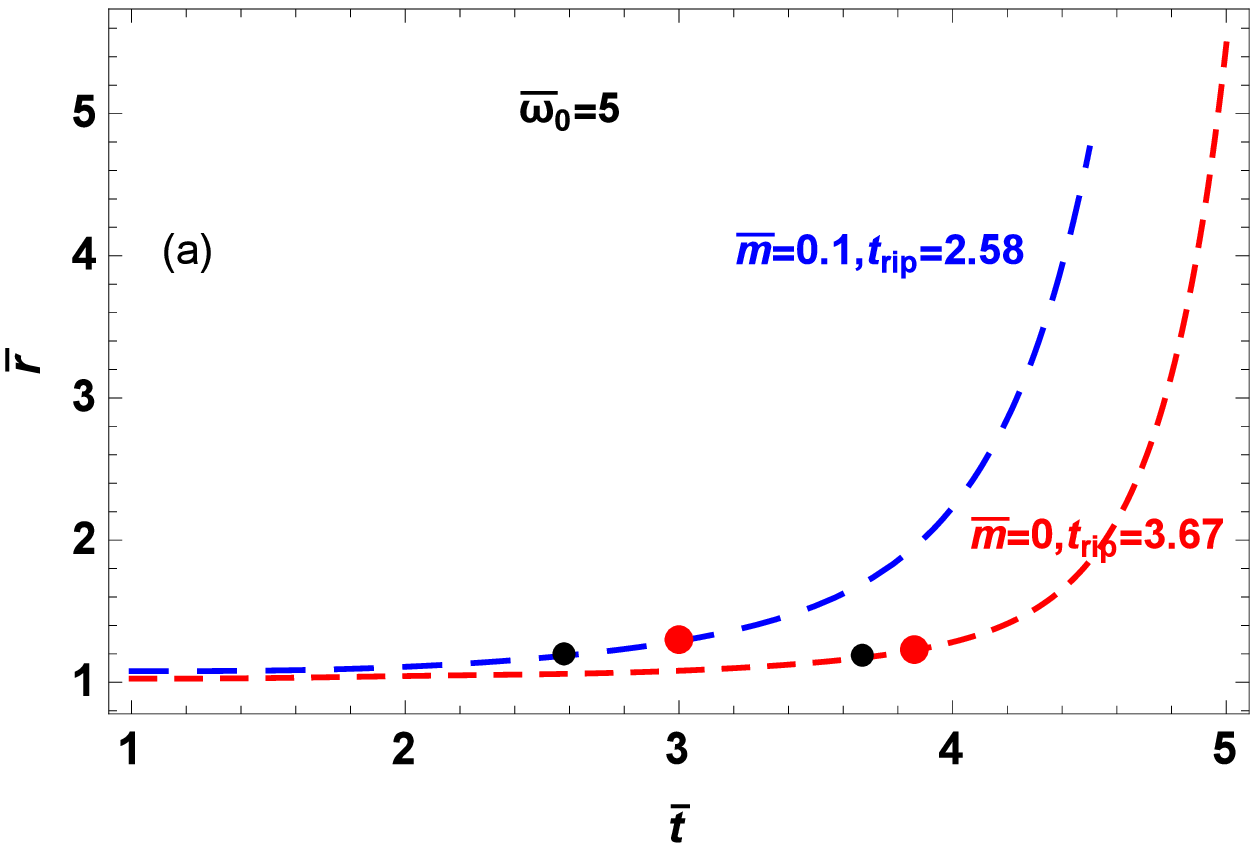}}}
\rotatebox{0}{\resizebox{0.48\textwidth}{!}{\includegraphics{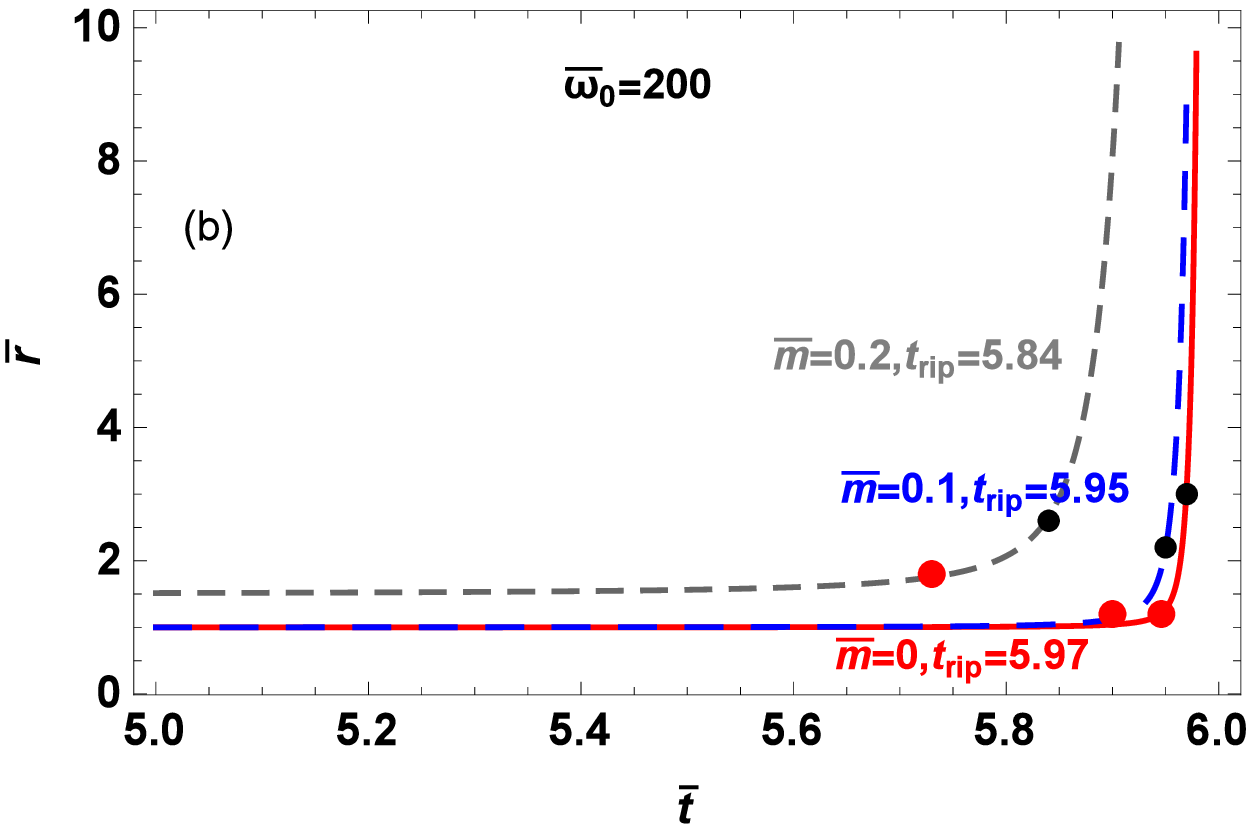}}}
\hspace{0pt}\caption{The radius $\bar r$ as a function of $\bar t$
when $\bar{\omega}_0=5$ and $\bar{\omega}_0=200$ for several
values of $\bar m$. The red points correspond to the time when the
size ${\bar r}(t)$ of the system has increased by about $20 \%$
compared to its equilibrium value. The black points correspond to
the time when the minimum of the effective potential disappears. Notice that scale of the $\bar t$ axis on the right plot is different and therefore the agreement between red and black points is much better.
}\vspace{1cm} \label{figbc}
\end{figure*}

At that time the system has been dissociated according to the full
relativistic analysis but it remains bound according to the
Newtonian approximation.

It is therefore clear that relativistic effects tend to
destabilize bound systems leading to an earlier dissociation
(smaller value of ${\bar t}_{rip}$) compared to the predictions in the
context of the Newtonian approximation. In the next section we
verify this result by a full numerical solution of the geodesic
equations (\ref{fradialeq}) and (\ref{ftimeeq}) and we present a
quantitative analysis of the magnitude of the relativistic
correction required for various bound systems defined by the
dimensionless parameters ${\bar \omega}_0$ and $\bar m$.

\section{ Quantitative Analysis: The time of bound system
Dissociation}

In the previous section we defined the time of dissociation of a
bound system as the time when the minimum of the effective
potential disappears due to the effects of the expansion. In the
context of a numerical solution of the system of geodesic
equations, this definition is not as useful because the effective
force and potential are only probed at the location of the
solution ${\bar r}({\bar t})$ with no information about neighboring values of
${\bar r}$ which could determine the binding status and stability of the
system.

\begin{table*}[t]
\centering \scalebox{1.2}{
\begin{tabular}{|c|c|c|c|c|c|}
  \hline
  \textbf{System}& \textbf{Mass($M_\odot$)} & \textbf{Size($Mpc$)} & \textbf{$\bar\omega_0$} &\textbf{$\bar m $} & \textbf{$\Delta {\bar t}_{rip}$} \\
  \hline
  Solar System & $1.0$ & $2.3\times10^{-9}$ & $3.5\times10^6$ & $2.1\times10^{-11}$ & $<10^{-8}$ \\
 Milky Way Galaxy & $1.0\times10^{12}$ & $1.7\times10^{-2}$ & $1.8\times10^2$ & $2.9\times10^{-6}$ & $2.4\times10^{-7}$ \\
  Typical Cluster & $1.0\times10^{15}$ & $1.0$ & $12$ & $4.9\times10^{-5}$ & $5.9\times10^{-5}$ \\
  Accretion Disk (neutron star) & $1.5$ & $3.3\times10^{-19}$ & $4.3\times10^{21}$ & $0.22$ & $<10^{-8}$ \\
  Hypothetical Large Massive & $3.0\times10^{20}$ & $1.0\times10^{2}$ & $9.1$ & $0.15$ & $0.93$ \\
  \hline
\end{tabular}
} \caption{The parameter values and the corresponding level of
relativistic corrections to the dissociation time for some typical
bound systems. The last column shows the difference in ${\bar
t}_{rip}$  between the Newtonian approximation and the
relativistic value  ${\bar t}_{nr_{rip}}-{\bar t}_{gr_{rip}}$ where
${\bar t}_{nr_{rip}}$ is the value of ${\bar t}_{rip}$ in the Newtonian
approximation and ${\bar t}_{gr_{rip}}$ the relativistic value.}
\label{table1}
\end{table*}

By comparing the dissociation times predicted by the effective
potential with the form of the trajectories ${\bar r}({\bar t})$ we concluded
that to within a good approximation the minimum of the effective
potential disappears when the solution ${\bar r}({\bar t})$ diverges by about
$20\%$ from its initial equilibrium value. We thus use this as a
criterion of dissociation when solving the system of geodesic
equations numerically. Due to the different nature of this
criterion we expect only qualitative agreement between the values
of ${\bar t}_{rip}$ obtained from the potential minimum and those
obtained from the numerical trajectories ${\bar r}({\bar t})$. However, as will
be discussed below in most cases the agreement is good even in the
quantitative level.

We solved the system of geodesic equations
(\ref{fradialeq})-(\ref{ftimeeq}) with initial conditions
corresponding to ${\bar t}_i =1$ and ${\bar r}_i$ corresponding to
the minimum of the effective potential at ${\bar t}={\bar t}_i=1$
(including expansion). This value was in all cases considered,
close to ${\bar r}=1$ corresponding to the minimum of the
effective potential without the effects of the expansion. In Fig.
4 we show the solution ${\bar r}(t)$ for ${\bar \omega}_0=5$,
${\bar \omega}_0=200$ when ${\bar m}=0.1$ superposed with the
corresponding radial function obtained in the Newtonian
approximation (${\bar m}=0$). The trend for earlier dissociation
in the relativistic treatment compared to the Newtonian approach
is clear. However, the difference of dissociation times decreases
as $\overline{\omega}_0$ increases.

As shown in Fig. 4 the bound system dissociation time ${\bar t}_{rip}$ is
well represented by the time when the size ${\bar r}(t)$ of the
system has increased by about $20 \%$ compared to its equilibrium
value. Given the rapid increase of the physical size of the system
after dissociation, the assumed relative size increase for
dissociation does affect significantly the obtained value for
${\bar t}_{rip}$. This is less accurate for larger systems
(smaller $\bar {\omega}_0$ shown in Fig. 4a) when the dissociation
proceeds more smoothly. Notice also that in all cases $\dot
{\bar r}$ is small before the dissociation which justifies the
fact that we ignored it in the construction of the effective
potential.

\begin{figure*}[!t]
\centering
\rotatebox{0}{\resizebox{0.48\textwidth}{!}{\includegraphics{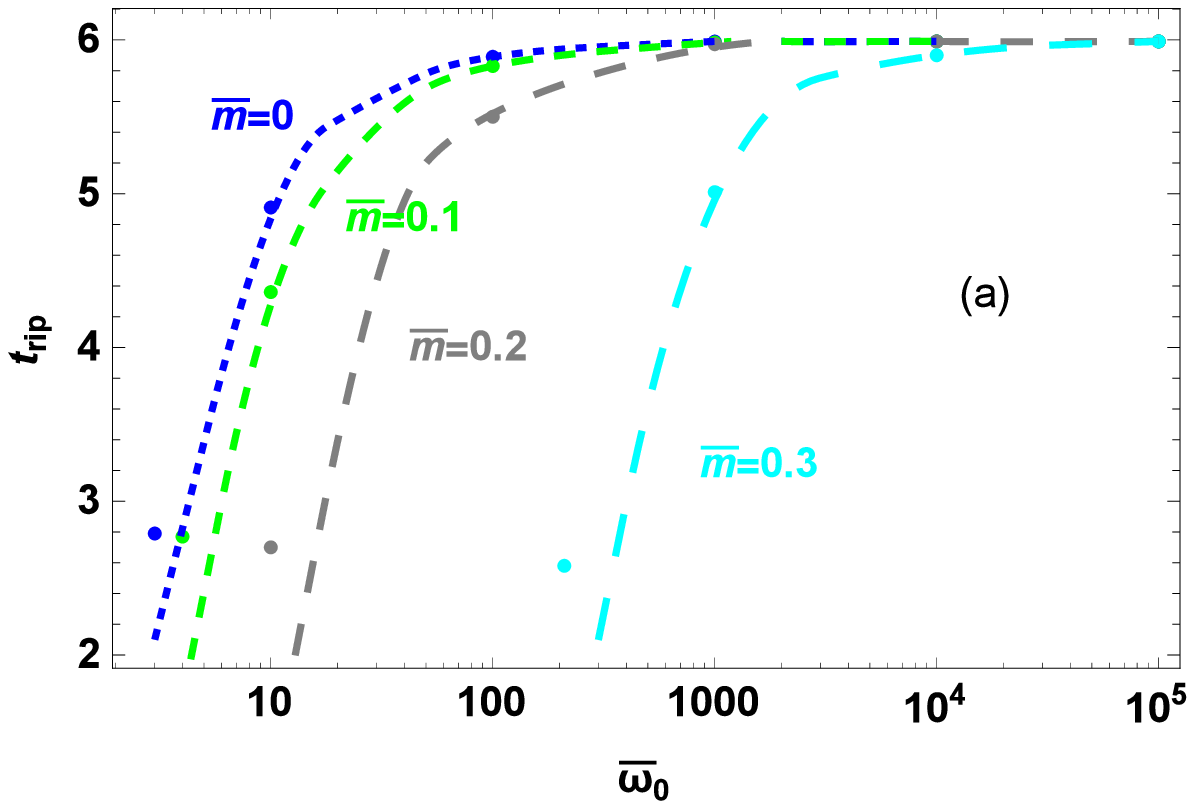}}}
\rotatebox{0}{\resizebox{0.48\textwidth}{!}{\includegraphics{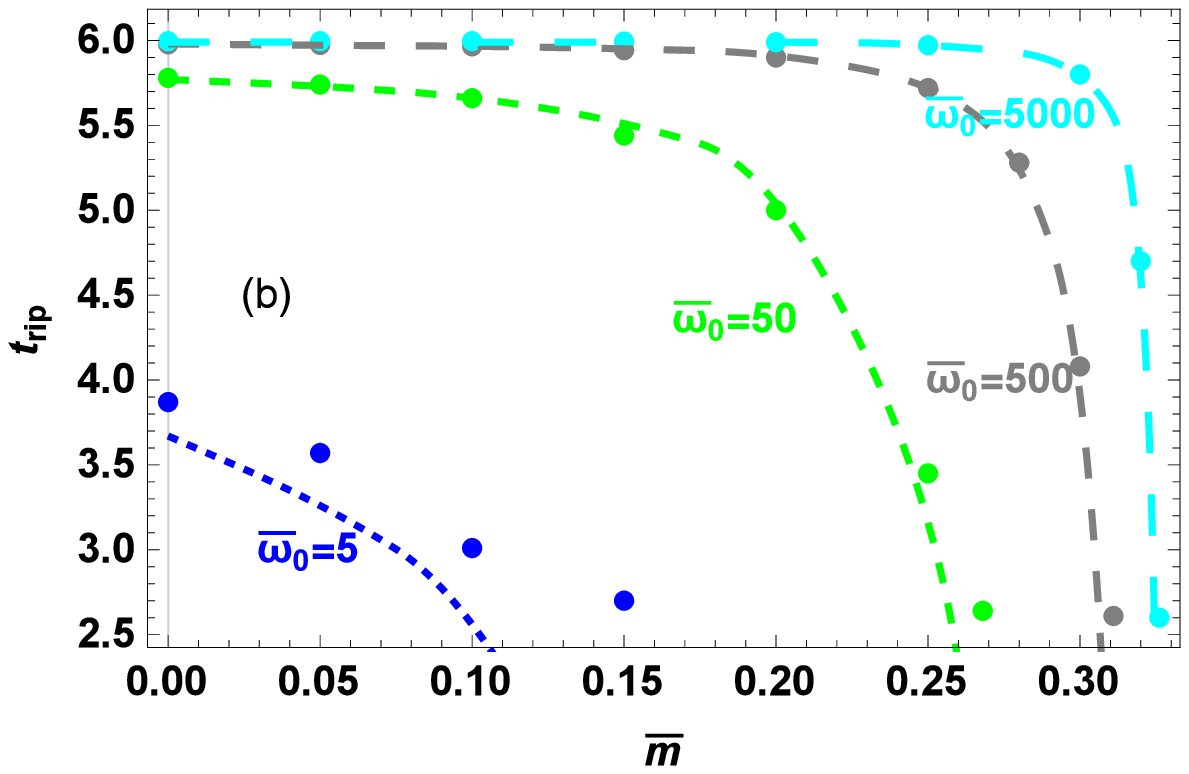}}}
\hspace{0pt}\caption{Figure 5a: The value of ${\bar t}_{rip}$ as a
function of ${\bar \omega}_0$ for various values of $\bar m$. Fig.
5b: The value of ${\bar t}_{rip}$ as a function of $\bar m$ for
various values of ${\bar \omega}_0$. The
thick dots correspond to dissociation times obtained using the numerical solution of the geodesic equations $r(t)$
while the lines were obtained using the effective potential of eq. (\ref{veff}) by finding the time when the potential
minimum disappears.}\vspace{1cm} \label{figd}
\end{figure*}

\begin{figure*}[!t]
\centering
\rotatebox{0}{\resizebox{0.6\textwidth}{!}{\includegraphics{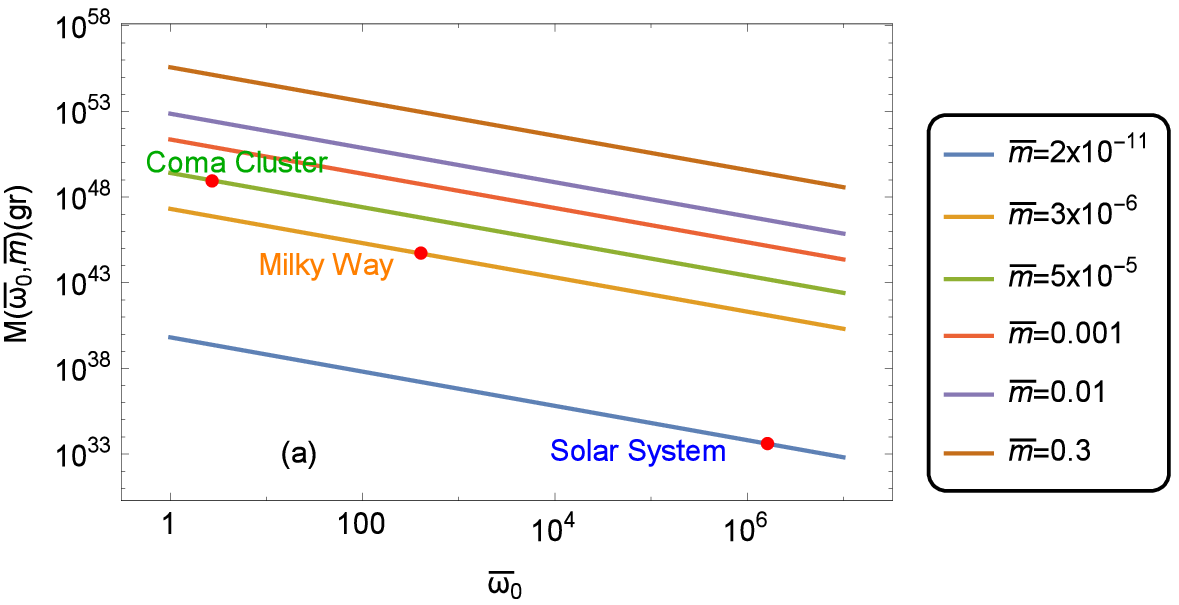}}}\\[5mm]
\rotatebox{0}{\resizebox{0.6\textwidth}{!}{\includegraphics{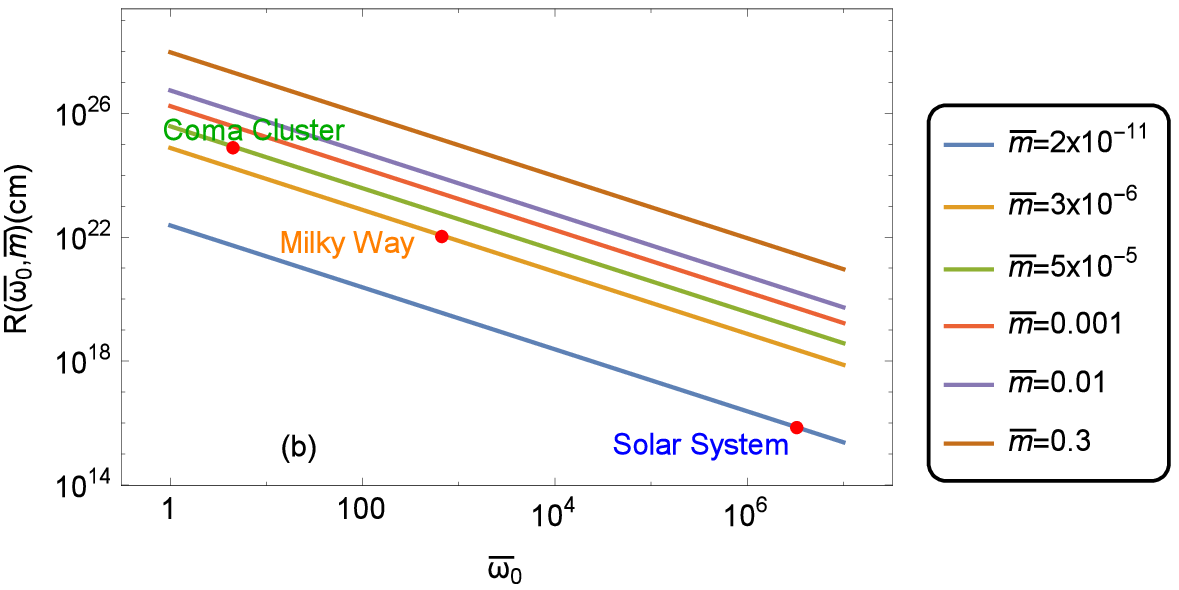}}}
\hspace{0pt}\caption{The mass of physical systems as a function of
the dimensionless parameter ${\bar \omega}_0$ for various values
of ${\bar m}$ (upper frame) and the size of physical systems in as
a function of the dimensionless parameter ${\bar \omega}_0$ for
various values of ${\bar m}$ (lower frame).}\vspace{1cm}
\label{fige}
\end{figure*}

Figure 5a shows the value of ${\bar t}_{rip}$ as a function of
${\bar \omega}_0$ for various values of $\bar m$. The curve for
$\bar m=0$ corresponds to the Newtonian limit. As
$\bar m$ increases, the relativistic correction to the value of
${\bar t}_{rip}$ increases dramatically for low values of $\bar
{\omega}_0$ (large massive systems). Therefore, the dissociation
of some large and strongly bound systems due to the expansion,
proceeds significantly earlier than anticipated in the context of
the Newtonian approach. This is also demonstrated in Fig. 5b where
we show  ${\bar t}_{rip}$ as a function of $\bar m$ for various
values of ${\bar \omega}_0$. The thick dots correspond to
dissociation times obtained using the numerical solution of the
geodesic equations ${\bar r}(t)$ while the lines were obtained
using the effective potential of eq. (\ref{veff}) by finding the
time when the potential minimum disappears.

\begin{table*}[t]
\centering \scalebox{1.2}{
\begin{tabular}{|c|c|c|c|c|c|}
  \hline
  \textbf{System}& \textbf{Mass($M_\odot$)} & \textbf{Size($Mpc$)} & \textbf{$\bar\omega_0$} &\textbf{$\bar m(\times10^{-5})$} & \textbf{${\Delta {\bar t}}_{rip}\times(10^{-5})$} \\
  \hline
  Typical Cluster & $10^{15}$ & $1.0$ & $12.4$ & $4.9$ & $5.9$ \\
  $ $ & $10^{15}$ & $0.8$ & $17.4$ & $6.1$ & $5.6$ \\
  $ $ & $10^{15}$ & $0.6$ & $26.8$ & $8.1$ & $4.6$ \\
  $ $ & $10^{15}$ & $0.4$ & $49.2$ & $12$ & $3.8$ \\
  $ $ & $10^{15}$ & $0.2$ & $139$ & $24$ & $2.7$ \\
  \hline
\end{tabular}
} \caption{The parameter values and the corresponding level of
relativistic corrections to the dissociation time for a typical
cluster, when we introduce a rescale in the size of the system. In the last column we have the difference of the Newtonian ${\bar t}_{nr-rip}$ minus the corresponding relativistic value. Notice that the relativistic rip occurs slightly earlier as expected but the difference from the Newtonian value decreases slowly with the rescaling to smaller sizes as the cosmological effects become less important.}
\label{table2}
\end{table*}

Notice however that systems with ${\bar \omega}_0$ larger than
about $10^4$ (relatively small systems) have dissociation times
${\bar t}_{rip}$ that are practically indistinguishable from the
Newtonian approximation independent of the value of $\bar m$. An
appreciable deviation of the value of ${\bar t}_{rip}$ from the Newtonian
approximation occurs for low values of ${\bar \omega}_0$ ($5-100$)
and large values of $\bar m$ ($O(10^{-1}))$. This range of
parameters corresponds to large and massive systems (eg size of
about 10-100Mpc and mass $10^6$ times larger than a typical
cluster of galaxies). Such systems where relativistic corrections are
important need to fulfil two conditions
\begin{enumerate}
    \item They need to be large so that the cosmological acceleration repulsive force
    to be important even at early times. Thus ${\bar t}_{rip}$ is relatively small
    (early dissociation) even at the Newtonian level allowing for significant change
    in the context of the relativistic correction.
    \item They also need to be massive so that their Schwarzschild radius(and the
     innermost stable orbit) to be comparable (a few times smaller) to their initial
     stable orbit radius.
\end{enumerate}

We stress that most cosmological bound systems have ${\bar m}$
which is much smaller than $\frac{1}{3}$. In particular for a
cluster of galaxies  ${\bar m}\simeq 10^{-5}$,  for a galaxy
${\bar m}\simeq 10^{-6}$ and for the solar system ${\bar m}\simeq
10^{-11}$. For such systems the Newtonian approach provides an
accurate approach for the dissociation time ${\bar t}_{rip}$.

Even some systems that are considered strongly bound (${\bar
m}\simeq 0.1$) such as  an accretion disk around a neutron star
are not large enough to have appreciable difference of ${\bar t}_{rip}$
due to relativistic effects (they have a very large ${\bar
\omega}_0$). A system with appreciable relativistic corrections of
the dissociation time would be  a hypothetical bound system with
mass $10^{20}$ $M_\odot$ and size about $100Mpc$ (about $10^6$
times more massive than a cluster of galaxies).

In Table I we show the parameter values and the corresponding
level of relativistic corrections to the dissociation time for
some typical bound systems

Fig. 6a shows the mass of physical systems as a function of the
dimensionless parameter ${\bar \omega}_0$ for various values of
${\bar m}$. Some physical bound systems are also indicated on the
plot. Similarly Fig. 6b shows the size of physical systems as a
function of the dimensionless parameter ${\bar \omega}_0$ for
various values of ${\bar m}$.  An accretion disk around a neutron
star ($r\simeq 50km$, $M\simeq 1.4M_{\odot}$) is out of the range
of these plots as it has ${\bar m} \simeq 0.1$ but
${\bar\omega}_0\simeq 10^{20}$ (see also eqs
(\ref{mbarvalue}),(\ref{om0value})). As shown in Table I, despite
the relative large value of $\bar m$ of such a strongly bound
system, its dissociation time would practically be identical to
the one derived in the context of the Newtonian approximation due
to its relatively small size and large value of ${\bar \omega}_0$.

Relativistic corrections tend to change slowly when the size of a
given bound decreases. Such a decrease implies an increase of both
${\bar m}$ and ${\bar \omega}_0$. The parameter values and the
corresponding relativistic corrections as the scale of a typical
cluster shrinks by a factor of 5 are shown in Table II. Notice
that the increase of ${\bar \omega}_0$ appears to be more
important during shrinking a system than the increase of ${\bar
m}$ and therefore the relativistic corrections to ${\bar t}_{rip}$
decrease slowly as the size of the bound system is reduced.

\section{ Conclusion-Discussion}
We have demonstrated that when relativistic effects are taken into
account, the dissociation of bound systems in phantom cosmologies
occurs earlier than it would have been predicted in the Newtonian
approximation used by previous studies. The correction in all
known bound systems is small. However, there are hypothetical
cosmologically large and massive bound systems where the
correction is significant.

Interesting extensions of the present analysis include the
following:
\begin{enumerate}
    \item The analysis of more general classes of geodesics like infalling
    radial geodesics with no angular momentum which at the time of the
     Big Rip are close or even beyond the black hole horizon.
    \item The use of McVittie geodesics to derive the  relativistic
    corrections on the turnaround radius which is the non-expanding shell
    furthest away from the center of a bound structure. In the context of
     the Newtonian approximation the maximum possible value of the turnaround
     radius for $w=-1$ ($\Lambda CDM$) is equal  to $(3GM/\Lambda c^2
     $) ~\cite{Pavlidou:2013zha}.
\end{enumerate}

{\bf Numerical Analysis Files:} Mathematica files that illustrate the implemented numerical analysis of the present study and the creation of some figures may be downloaded from \href{https://dl.dropboxusercontent.com/u/20653799/pub-math/mcvittie-geod-math.rar}{this url}

\end{document}